# Hiding levitating objects above a ground plane

Jingjing Zhang[1*], Yu Luo[2], and Niels Asger Mortensen[1]

[1] *DTU Fotonik - Department of Photonics Engineering, Technical University of Denmark, DK-2800 Kongens Lyngby, Denmark*

[2] *The Blackett Laboratory, Department of Physics, Imperial College London, London SW7 2AZ, UK*

**Abstract**

An approach to hiding objects levitating or flying above a conducting sheet is suggested in this letter. The proposed device makes use of isotropic negative-refractive-index materials without extreme material parameters, and creates an illusion of a remote conducting sheet. Numerical simulations are performed to investigate the performance of this cloak in two-dimensional (2D) and three-dimensional (3D) cases.

[*] *Author to whom correspondence should be addressed; electronic mail: jinz@fotonik.dtu.dk*

Recently, Transformation optics has inspired the idea of realizing electromagnetic (EM) transparency by forcing the waves to bend around arbitrarily sized and shaped volumes [1-3]. These theoretical considerations were later experimentally validated in microwave frequencies [4], using concentric arrays of split ring resonators. This approach, however, can only be achieved at a single frequency or very narrow band due to the strong anisotropy at the inner boundary. In order to broaden the working frequency range for the cloak, other methods which avoid resonant elements have been proposed [5–7]. And the so called "carpet cloak" [5], which can effectively camouflage a bump on the ground plane over a broad band of frequency range, has already been experimentally realized in 2D and 3D scenarios [8-11].

Many other schemes are also considered to realize invisibility for different application situations. One route uses plasmonic shells with a suitable design that can compensate the scattering from a given object [12], and can be potentially extended for multiple frequencies utilizing multiple shells around arrayed objects [13]. Cloaking of polarizable dipoles and line quadrupoles in the vicinity of a flat or cylindrical superlens are also studied [14, 15], based on anomalous localized resonance. Transmission-line networks have been applied to cloak an array of metallic objects [16] and experimental validation has subsequently been reported [17].

The previous designs of the cloak with transformation approach mostly rely on making its contents appear small enough to be invisible. However, for EM wave detections, cloaking of an object does not necessarily require making the object invisible. In this Letter, we suggest a device which aims to camouflage an object by creating an illusion of a remote PEC sheet. This idea is in some sense similar to that of carpet cloak [5]. However, compared with the carpet cloak which must necessarily touch the ground plane, the cloak proposed here is levitating in the midair above the PEC ground. Hence, it can potentially hide levitating or flying objects far away from the ground. The cloak is constructed with isotropic negative-refractive-index metamaterials without extreme material parameters. To further illustrate this idea, both the 2D and 3D cases are studied with numerical simulations.

Our approach takes advantage of the principle that a perfect conducting plane is inherently invisible to a specifically polarized plane wave when the electric field is perpendicular to this conducing plane, as shown in Fig. 1(a). Similarly, for a curved conducting sheet, invisibility can still be achieved by guiding the incident wave and keeping the electric field perpendicular to the curved surface [see Fig. 1(b)]. Consequently, any object can be hidden underneath the bump without being detected, which is similar to the idea of the carpet cloak [5, 8-11]. Next, if a device can guide the impinging wave around a closed volume and keep the electric field normal to its conducting surface everywhere, as depicted in Fig. 1(c), the obstacles inside this closed volume would be invisible to the incident wave. A special example which can achieve this effect shown in Fig. 1(c) is the recently proposed super-concentrator or super-scatterer device [18-21]. As depicted in Fig. 1(d), this device is achieved by squeezing a cylindrical (or spherical) region (with a radius $R_2^2/R_1$) denoted by the black dot line into the smaller concentric gray region (with a radius $R_1$). The detailed illustrations are given in [18-21]. In what follows, we make use of this transformation to achieve the concealing of levitating obstacles and start our discussion with the 2D scenario. For TM wave incidence, the resulting parameters in the physical space are given as

$$\begin{aligned}&\mu=\left(R_2/R_1\right)^4,\varepsilon=1 \text{ when } r<R_1\\&\mu=-\left(R_2/r\right)^4,\varepsilon=-1 \text{ when } R_1<r<R_2\end{aligned}. \quad (1)$$

Eq. (1) shows an isotropic property with no extreme parameter values, hence can be physically realized with metamterials [22, 23]. However, due to the intrinsic dispersive property of metamaterials, this cloak has a limited working frequency range as compared with the carpet cloak.

The ray trajectories of light passing through the whole device are also shown in Fig. 1(d). The blue solid lines denote the rays propagating in the space where some of them are guided through the cloak, penetrating the brown region without scattering. Significantly, one can notice that closed loops of rays are formed inside the medium (purple solid lines). As demonstrated in Ref. [19, 20], these circulating loops are coupled to some rays within the region $R_2 < r < R_2^2 / R_1$ that do not directly impinge on the medium (purple dashed line) by evanescent waves. Since the power that flows through the brown core is larger than that flowing through the whole system, this kind of device has also been used to implement a concentrator [18]. The coating ($R_1 < r < R_2$) composed of negative-index-material has the similar function as a superlens [24], hence has been applied to design the super scatterer [21] and the complimentary cloak [25]. As illustrated above, for TM wave incidence, one can put a PEC surface along the light loops denoted by purple solid lines without being detected. In addition, it has been demonstrated in Ref. [19] that the purple solid loops inside the device in the physical space are mapped to the purple dashed lines outside the device in the virtual space. Therefore, for outer wave detection, this closed PEC surface will behave exactly the same as a PEC plane outside the device (one of the purple dashed lines), and would be invisible for TM waves whose electric field is perpendicular to this plane.

We use a finite element solver (COMSOL Multiphysics) to verify this idea. The operating frequency of the plane wave source is 2GHz and $R_1 = 0.1m$, $R_2 = 0.2m$. In Fig. 2 (a), a TM polarized plane wave is launched in the *x* direction upon the device. As expected, no scattering is caused, showing that the PEC interior boundaries of the cloak are not sensed by the outer wave. For comparison, the field distribution for the illusion of PEC planes created by the system, which is $0.1m$ away from the outer boundary of the system, is denoted in Fig. 2 (b). In Fig. 2 (c) and (d), we study the field distributions for the system and the corresponding virtual PEC plane, respectively, when the EM wave is incident in *y* direction, and the agreement between the two panels further demonstrates our analysis.

Although total transparency in the free space only happens for a specially polarized wave along a certain direction, this device can make a cloaked object undetectable for any incident detection in an environment with reflecting ground, because any cloaked obstacle can be disguised as the ground, similar to the idea of the carpet cloak. Here we emphasize that a carpet cloak only works when sitting on the conducting plane and is not suitable for the case where an object is levitating above the ground at a certain height. However, as may be noticed from Fig. 2 (b), the virtual PEC illusion created by this system can be some distance away from the outer boundary of the proposed camouflage cloak, indicating that there is space between the whole device and the ground. Fig. 3 (a) depicts the field distribution when a 2GHz Gaussian beam is incident upon a PEC ground plane. In Panel (b), a cloaked PEC scatterer is placed above this ground. As long as it is located at the appropriate height, the total field distribution will resemble that in Panel (a) because the PEC plate that the device mimics will correctly superpose the ground plane. As a comparison, we also simulate the bare PEC obstacle which will cause pronounced scattering, as shown in Panel (c). Although TM illumination is used in the simulation, similar results can also be observed in a TE incidence case. The above results validate the capability of camouflaging objects levitating in the midair. This property may open a new perspective for hiding objects above different kinds of practical reflecting surfaces, for instance, the water surface and the smooth ground, as long as the total reflection condition is satisfied. It is also worth mentioning that since the mimicked PEC plane does not touch the cloaked obstacle, the observer is still unable to find the accurate position of the real target even if the virtual PEC plane is detected.

For 3D cases, the material parameters of the camouflage cloak are given as

$$\begin{aligned}\varepsilon = \mu = \left(R_2/R_1\right)^2 \quad \text{when } r < R_1 \\ \varepsilon = \mu = -\left(R_2/r\right)^2 \quad \text{when } R_1 < r < R_2\end{aligned} \qquad (2)$$

Similarly, the 3D cloak still preserves the advantage that the material parameters have no extreme value. Fig. 4 shows the magnetic field distribution for a spherical camouflage device under a plane wave illumination [panel (a)] and a Gaussian wave illumination [panel (b)]. In panel (b), the coated obstacle is 0.1m away from the conducting ground denoted by the red plane, and the reflection from the coated obstacle perfectly resembles that from the ground plane. In real applications when a coating with streamline contour is required, the design can be similarly treated in prolate spheroidal or ellipsoidal coordinate systems [26].

In conclusion, we propose a camouflage device which conceals an object by creating an image of a remote PEC plane. This property opens up interesting possibilities, such as hiding a levitating object above the reflective surface without touching the ground plane, and works for all the directions of incident waves. The device is realizable using isotropic metamaterials with spatially variable permittivity/permeability.

The work is supported by the Hans Christian Ørsted postdoctoral fellowship.

Figure Captions

Fig. 1 (color online) (a) A plane wave propagating along a PEC plane, with the electric field perpendicular to this plane. (b) A plane wave propagating along a curved PEC surface, with the electric field perpendicular to this surface. (c) A wave propagating along a closed PEC sheet, with the electric field perpendicular to this surface. The brown region indicates the transformation medium which renders the wave loop. (d) A specific example of the cloak which gives rise to a closed ray loop. Ray trajectories are shown when the cloak is under a plane wave illumination.

Fig. 2 (color online) (2D) (a) Magnetic field distributions for a cloak under a TM plane wave illumination in the *x* direction. (b) A PEC plane at $y = -0.3m$ under a TM plane wave illumination in the *x* direction. (c) Magnetic field distributions for a cloak under a TM plane wave illumination in the *y* direction. (d) A PEC plane at $y = -0.3m$ under a TM plane wave illumination in the *y* direction.

Fig. 3 (color online) (2D) (a) Magnetic field distribution when a Gaussian beam is launched at 45° towards the ground plane (a conducting sheet) from the left. (b) Magnetic field distribution when a PEC scatterer surrounded by a camouflage cloak is present. (c) Magnetic field distribution when a PEC scatterer is present without the cloak.

Fig. 4 (color online) (3D) (a) Magnetic field distributions for a spherical cloak under a plane wave illumination in the *x* direction. (b) Magnetic field distribution when a Gaussian beam is launched at 45° towards the ground plane from the left. A PEC scatterer surrounded by a camouflage cloak is present above this plane.

Fig. 1

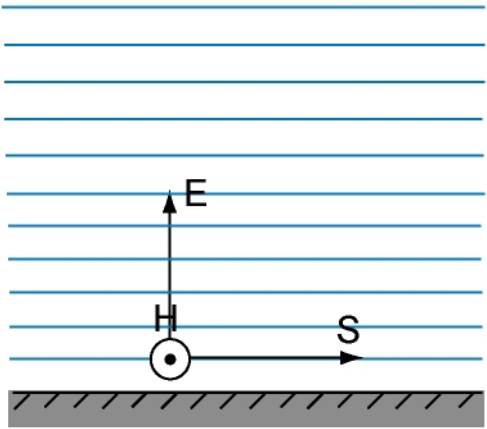
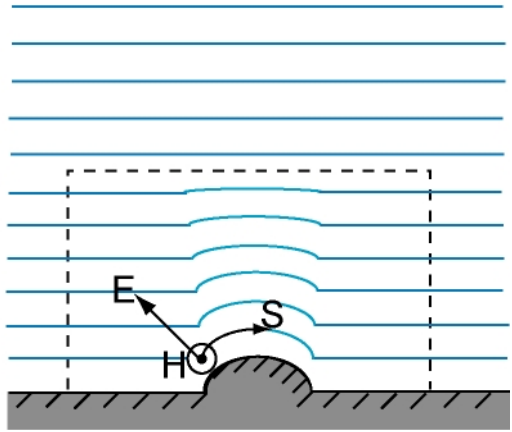
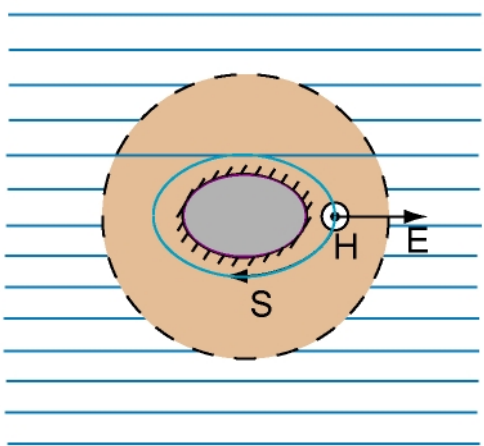
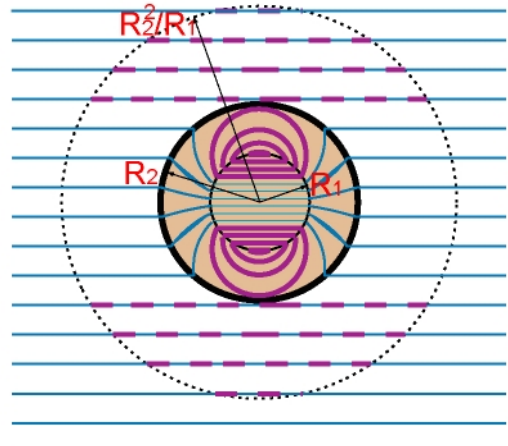

Fig. 2

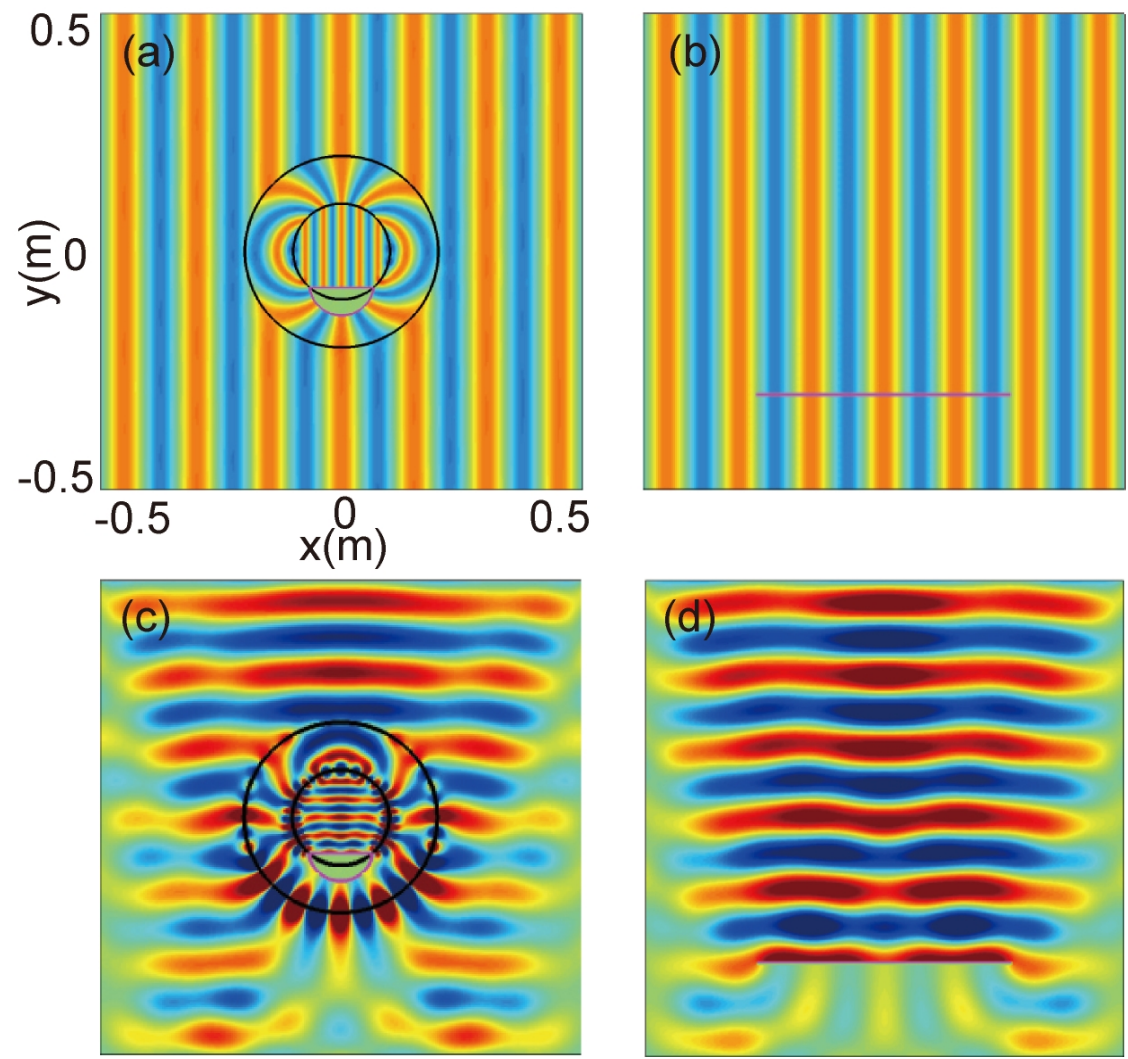

Fig. 3

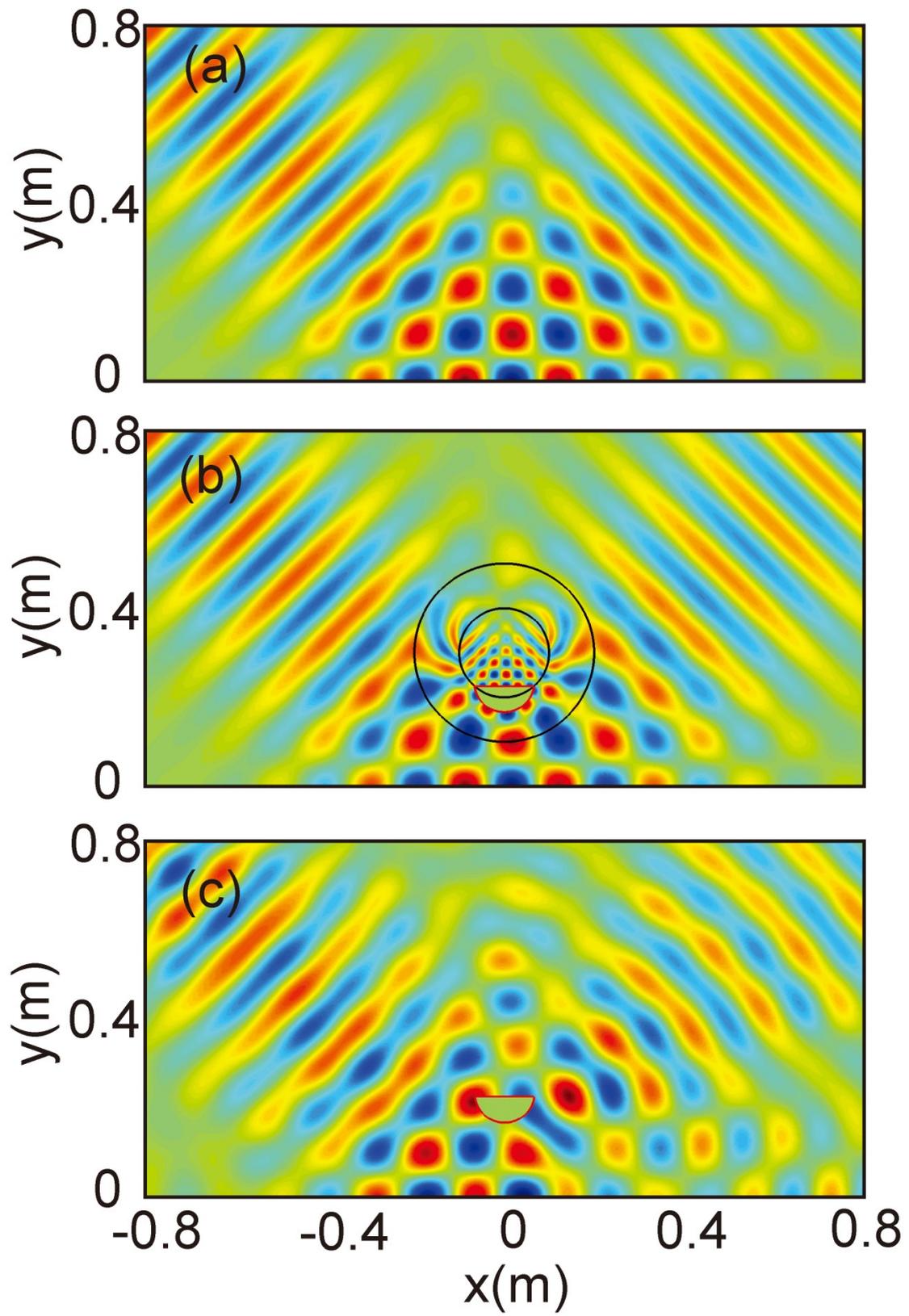

Fig. 4

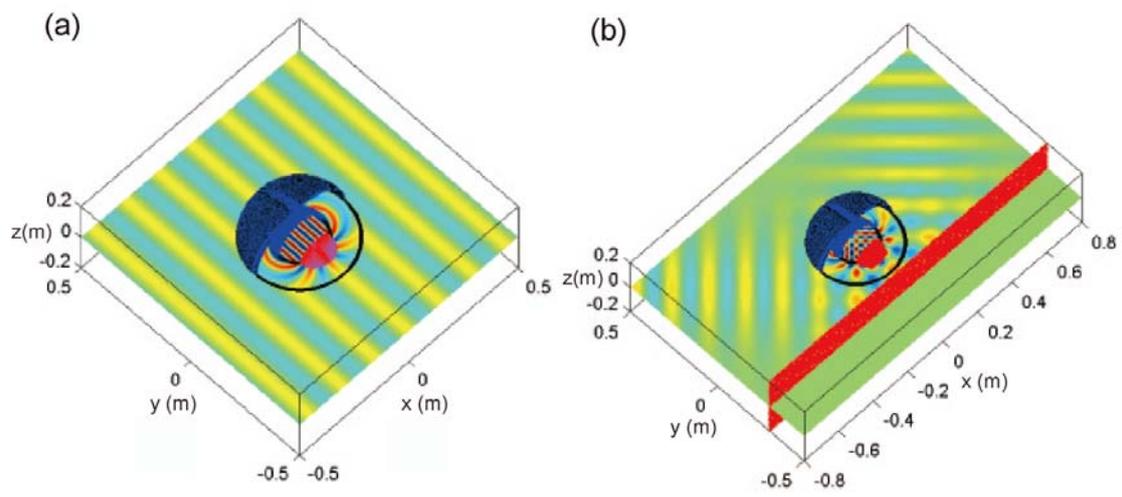